\documentclass[a4paper]{article}
\usepackage{anyfontsize}
\usepackage{INTERSPEECH2021}
\usepackage[table]{xcolor}
\usepackage{booktabs}
\usepackage{graphicx}
\usepackage[noadjust]{cite}
    
\usepackage{hyperref}
\usepackage{caption}
\usepackage{amsmath}
\usepackage{graphicx} % \scalebox
\usepackage{environ}
\usepackage{bm}
\usepackage{float}
\usepackage{pifont}% http://ctan.org/pkg/pifont
\newcommand{\cmark}{\ding{51}}%
\newcommand{\xmark}{\ding{55}}%
\title{  ConvRNN-T: Convolutional Augmented Recurrent Neural Network Transducers for Streaming Speech Recognition }
\name{Martin Radfar, Rohit  Barnwal, Rupak Vignesh Swaminathan, Feng-Ju Chang, Grant P. Strimel, Nathan Susanj, Athanasios Mouchtaris  }
\address{Alexa Machine Learning, Amazon, USA }
\email{\{radfarmr,rbarnwal,swarupak,fengjc,gsstrime,nsusanj,mouchta\}@amazon.com}

\begin{document}

\maketitle
\begin{abstract}
The recurrent neural network  transducer (RNN-T) is a prominent streaming end-to-end (E2E)  ASR technology. In RNN-T, the acoustic encoder commonly consists of stacks of LSTMs. Very recently, as an alternative to LSTM layers, the Conformer architecture was introduced where the encoder of RNN-T is replaced with a modified Transformer encoder composed of convolutional layers at the frontend and between attention layers. In this paper, we introduce a new  streaming ASR model, Convolutional Augmented Recurrent Neural Network Transducers (ConvRNN-T) in which we augment the LSTM-based RNN-T with  a novel convolutional frontend consisting of  local and  global context CNN  encoders. ConvRNN-T  takes advantage of causal 1-D convolutional layers, squeeze-and-excitation,  dilation, and residual blocks to provide both global and local audio context representation to LSTM layers. We show ConvRNN-T outperforms  RNN-T, Conformer,  and ContextNet on Librispeech  and in-house data. In addition, ConvRNN-T  offers less computational complexity compared to  Conformer.  ConvRNN-T's superior accuracy along with its low footprint make it a promising candidate for  on-device streaming ASR technologies.
\end{abstract}

\noindent\textbf{Index Terms}: RNN-T, streaming Automatic speech recognition (ASR), Convolutional neural networks, sequence-to-sequence, Transformer

\section{Introduction}
Sequence-to-Sequence (Seq2Seq) neural based ASR models  have revolutionized the ASR technology as they can obviate the need for text-audio alignment,   remove prohibitive HMM training, allow end-to-end training,  are streamable, and offer better accuracy with lower footprint  \cite{graves2012sequence,chan2015listen,chorowski2015attention,he2019streaming}. RNN-T \cite{graves2012sequence} is the most popular  streaming ASR model which has achieved lower word error rate (WER) compared to its predecessor, CTC-based ASR \cite{graves2006connectionist}.  RNN-T deploys a powerful loss function in which the posterior probability of output tokens is computed over all  possible alignments, a novel strategy that elegantly leverages the forward-backward algorithm \cite{Rabiner89-ATO}.  In addition, RNN-T employs a label encoder which implicitly incorporates a language model that auto-regressively predicts output tokens given latent audio representations and previous decoded tokens in a causal manner. 

With the emergence of  Transformers \cite{vaswani2017attention} and their outstanding success in language modeling,  Transformers  are deployed as neural acoustic encoders in Seq2Seq ASR architectures \cite{dong2018speech,moritz2020streaming}. The pioneering Transformer models are, however, non-streamable as they present the entire audio context to the decoder, and  use non-causal attention \cite{dong2018speech}. In order to make the Transformer streamable, Transformer-Transducers are introduced \cite{zhang2020transformer,moritz2020streaming,yeh2019transformer,tripathi2020transformer,chang2021multi,huang2020conv,chang2021context} in which the encoder of RNN-T is replaced with a Transformer encoder with  causal attention which means  the current audio frame only attends to left context.  There are also other mainstream of  attention based end-to-end ASR models that use an additive attention at the encoder output \cite{chan2015listen,chorowski2015attention}. To make these models streamable, monotonic attention and its variants  are proposed \cite{chiu2017monotonic,hsiao2020online}. One of the drawbacks of  attention based models is, however, their quadratic computational complexity which makes them not attractive for on-device applications.

Convolutional neural networks are another popular architecture for acoustic modeling in ASR systems \cite{abdel2014convolutional,han2020contextnet,li2019jasper,kriman2020quartznet}. Convolutional kernels  learn efficiently  the correlation between audio frames and provide high level latent representation of audio contexts that can better render to text. Inspired by recent advances in audio synthesis using causal dilated temporal convolutional neural networks \cite{oord2016wavenet} and outstanding performance improvement in CNN-based computer vision using the squeeze-and-excitation (SE) approach \cite{hu2018squeeze}, recently several convolutional based acoustic encoders have been introduced for Seq2Seq ASR models \cite{han2020contextnet,li2019jasper,kriman2020quartznet}. Among these models, ContextNet exhibits the state-of-the-art results comparable with those of Transformers and RNN based approaches\cite{han2020contextnet}. Convolutional based models, similar to Transformers, enable to perform  parallel processing but  are less computationally expensive. They, however, need a large number of layers to learn complex audio context. In addition, compared to recurrent models, CNNs need more memory consumption during inference  \cite{bai2018empirical}.

In order to improve the performance of Transformer transducers and make Transformers less dependent on positional encoding, Transformers with a convolutional frontend are proposed \cite{yeh2019transformer,mohamed2019transformers,li2020better}. In these models, often two to four layers of CNN are stacked with batch normalization, non-linear activation, and possible downsampling.  Using this strategy and given improvements observed for ContextNet \cite{han2020contextnet}, the Conformer model was proposed in which the convolutional blocks are not only used as the frontend but also inserted between multi-head self-attention blocks and feedforward networks in the encoder block \cite{gulati2020conformer}.  Introducing convolutional blocks into Transformer improves the performance substantially  and makes Conformer one of the best E2E  ASR models. 

 The combination of CNN and LSTM has been previously shown to improve the performance of both speech recognition  \cite{7953077} and speech enhancement models \cite{180500579}. In this work, instead of using an all CNN encoder, like ContextNet,  or combining multi-head attention with CNN, like Conformer in the RNNT architecture, we propose a new effective neural  E2E ASR transducer architecture called ConvRNN-T which augments LSTM-based RNN-T with a  convolutional frontend  which significantly improves the performance of RNN-T. ConvRNN-T takes advantage of both CNN and LSTM and prevents computational burden of multi-head attention. ConvRNN-T architecture is inspired by  (a): recent advances of temporal convolutional modules (TCM)  applied for speech enhancement \cite{pandey2019tcnn}, (b): causal separable dilated CNNs used in audio synthesis to provide a large receptive field for CNN \cite{oord2016wavenet}, (c): the squeeze-and-excitation layer proposed in \cite{han2020contextnet,hu2018squeeze } to allow to incorporate global context, and  (d): the Conformer convolutional block \cite{gulati2020conformer}. ConvRNN-T uses two convolutional blocks that encode local and global context for LSTM layers. We benchmark ConvRNN-T on Librispeech and on our de-identified in-house data.  Our experiments show ConvRNN-T significantly reduces WER when compared with RNN-T, ContextNet, and Conformer. ConvRNN-T is an attractive candidate for on-device applications when low computational complexity is as important as accuracy.

 \begin{figure}[h]
  \centering
  \includegraphics[width=.6\linewidth]{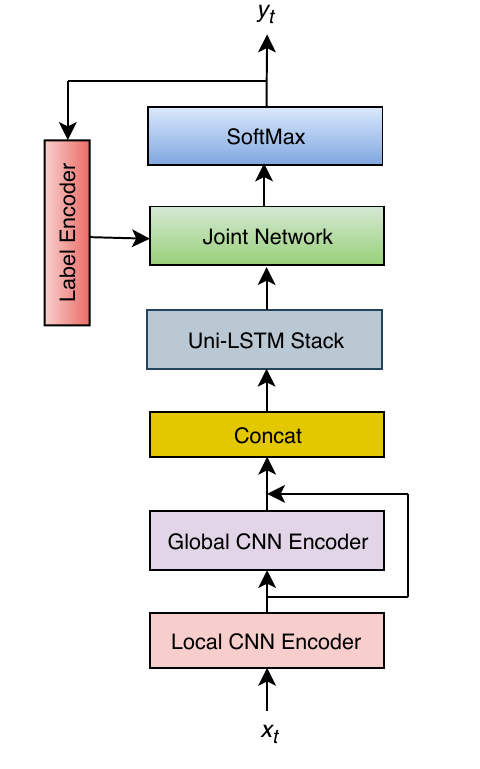}
  \caption{ A high-level block diagram of ConvRNN-T.}
  \label{fig:fans}
\end{figure}

\section{RNN-T Basics}

The input audio signal is transferred to a time-frequency  space represented by  $\bm{x}=[x_1,\ldots, x_i,\ldots,x_T]^\mathsf{T}$  where $x_i \in \mathcal{R}^D$, $T$ and $D$ denote the number of frames and the dimension of the frequency space, respectively;  also $\mathsf{T}$ denotes the transpose operation. Each audio signal, $\bm{x}$, has a transcript  represented by sequence of labels  $\hat{\bm{y}}=[\hat{y_1},\ldots, \hat{y_j},\ldots,\hat{y_L}]^\mathsf{T}$, of the length $L$. The labels can be phonemes, graphemes, or word-pieces. In E2E ASR transducer models, the goal is to find the model parameters (i.e., all learnable weights  in the model) that maximize the posterior probability of  alignment between audio and text based on pairs of audio-transcript training samples. RNN-T uses a null token that indicates the lack of token for a time step. To compute the loss,  RNN-T uses a trellis whose $X$ and $Y$ coordinates are time steps and  tokens, respectively. On this trellis,  a move upward  indicates to emit a token and a move  forward indicates to move to the next time step without emitting any token.  For each node on this trellis, we compute the posterior probability of alignment until we reach the up-right corner of the trellis (i.e. $(T,L)$ node).  To make the loss---the negative log of the posterior probability of alignment--- differentiable and tractable, \cite{graves2012sequence} proposed to use the forward-backward algorithm.  RNN-T conditions the posterior probability of the current token on the all previous predicted tokens as opposed to the CTC model. RNN-T leverages a label encoder, consisting of uniLSTM layers, which implicitly incorporates  a language model to predict the current latent label based on previous predicted tokens. The  details of how to compute the loss can be found in \cite{graves2012sequence}.
\section{ConvRNN-T Architecture}
 Figure 1 illustrates a high-level block diagram of ConvRNN-T. The architecture consists of a uniLSTM RNN-T and local and global CNN-based encoders. The input features are  fed to the local CNN-based encoder; the output of which is then fed to the global CNN based encoder; next, the uniLSTM encoder takes as input the concatenated outputs of the  two CNN blocks after they are projected to a smaller subspace. The outputs of the uniLSTM stack  are combined with the outputs of the label encoder  given the previous non-blank label sequence and fed to SoftMax  to compute the posterior probability  of the token. The label encoder is a uniLSTM whose input is previous predicted tokens. In the following subsections, we elaborate on local and global CNN encoders.

 \begin{figure}[h]
  \centering
  \includegraphics[width=.5\linewidth]{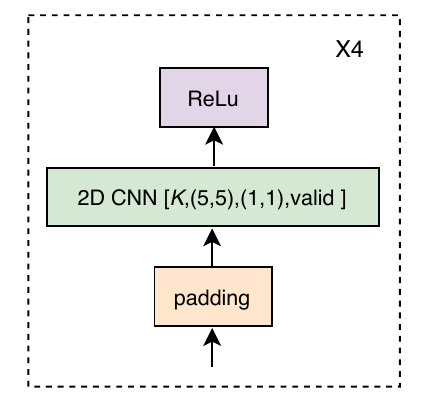}
  \caption{ The local CNN encoder block. The numbers in bracket are  associated with the size of filter (K is set to 100 and 64 for the first two and last two convolutional layers), kernel size, stride and  padding which is set to be valid---this is for making convolution causal. The padding block is added to make the length of input and output equal.}
  \label{fig:fans}
\end{figure}

\begin{figure*}[t]
  \centering
  \includegraphics[width=.8\linewidth]{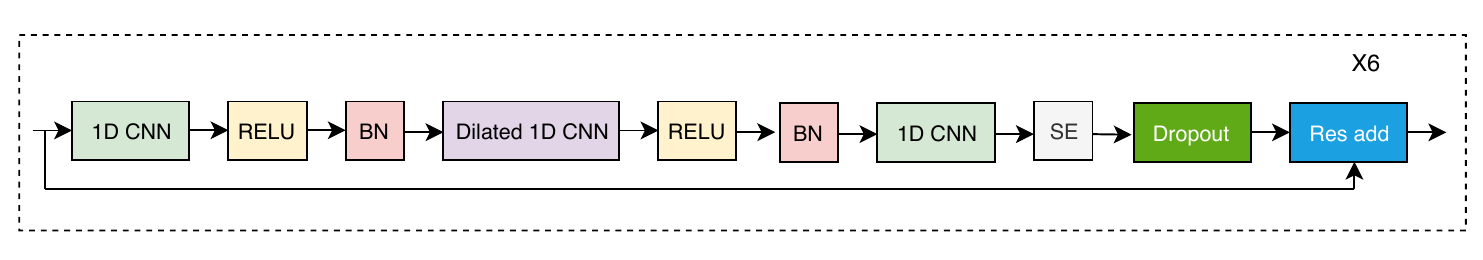}
  \caption{The structure of the global CNN encoder, consisting of three layers of 1-D causal CNNs, two of which are point-wise and one is dilated depth-wise. The squeeze-and-excitation (SE) approach is applied after the third CNN layer.}
  \label{fig:slutask}
\end{figure*}

%\begin{figure}[h]
  %\centering
  %\includegraphics[width=.7\linewidth]{Fig4.png}
  %\caption{The process of building causal convolution by zero padding for a 1D toy example.}
  %\label{fig:fans}
%\end{figure}

 \subsection{Local CNN encoder} 
As depicted in Figure 2, the first convolutional block consists of four layers of 2-D CNN. Each CNN is followed by a ReLu activation. Because our model is streamable, we need to ensure that CNN computations are causal. A system is causal when its output at time $t_0$ only depends on input at $t\le t_0$. In order to make CNN causal, we zero-pad the time dimension from left with kernel size minus one zeros. For ConvRNN-T, we used the kernel size of five and  stride of size one. Also, we use 100 channels for the first two CNN layers and 64 for the last two layers. This strategy of using layers of CNN  as a frontend has been tried in previous Transformer based transducers and has shown improvements \cite{yeh2019transformer,mohamed2019transformers}
  \subsection{Global CNN encoder } 
The global CNN encoder is inspired by the TCM block in the convolutional architecture that applied to the problem of speech enhancement \cite{pandey2019tcnn} as well as the residual block in ContextNet \cite{han2020contextnet,zagoruyko2016wide}. The global CNN encoder uses two layers of point-wise 1-D CNN, and one depth-wise 1-D CNN which is sandwiched between the two point-wise 1-D CNN layers. Here, we only use convolutional kernel on time steps and consider the features as channels. The first two 1-D CNNs are followed by ReLu activation and batch normalization. We also add a residual connection which has been shown to alleviate vanishing gradient and speed up convergence \cite{zagoruyko2016wide}.  Using depth-wise and point-wise convolution reduces the computational complexity significantly and has shown to give better results in large scale vision problems \cite{chollet2017xception}.

There are two known problems with CNNs. First, when CNNs are used to model  time series, they do not have access to long range contextual information as opposed to Transformers which use attention to collect information from past and future frames.  Second,  CNN-based models need many layers or large kernels to increase their receptive fields.  The squeeze-and-excitation  (SE) approach \cite{hu2018squeeze} and dilated kernels \cite{oord2016wavenet} are suggested to mitigate these issues. A dilated CNN expands  the filter to cover  an area larger than its length by skipping input values with a certain step. 
 In the SE approach, we compute the mean over all previous time steps and apply two non-linear transformations to the mean vector. We call this vector as the $\textit{context vector}$. We then  multiply the current output by the context vector in an element-wise manner. This way, we incorporate  global information from all previous frames to the current time step. Mathematically, the SE approach can be summarized as follows: 
 $z^{\text{SE}}_i=z_i\otimes \left (\text{Sigmoid}(W_1(\text{ReLu}(W_2\times \frac{1}{T_i}\sum_{k=1}^{T_i} z_k))) \right) $
 where $z_i$ is the output of last point-wise 1D CNN for the time step $i$; $\otimes$ denotes the element-wise multiplication, $W_1$ and $W_2$ are learnable weight matrices. At the last step, we apply dropout as a regularization method before we  add the residual block. The global encoder consists of  six blocks of the architecture as depicted  in Figure 3. The final output of the global encoder is concatenated with the output of the local encoder and transformed to the original input dimension. The resulting tensor is used as the input to the stacks of uniLSTMs.

 \section{Experiments}
\subsection{Data and Model parameters}

We use the Librispeech corpus which consists of 970 hours of labeled speech  \cite{panayotov2015librispeech}  and  50K hours of our de-identified in-house data  to benchmark ConvRNN-T against RNN-T, Conformer, and ContextNet.  We used 64-dimensional log short time Fourier transform  vectors obtained by segmenting the utterances with a Hamming window of the length 25 ms and frame rate of 10 ms. The 3 frames are stacked with a skip rate of 3, resulting  in 192-dimensional input features. These features are normalized using the global mean and variance normalization.
 \begin{table}[t]
 \small
  \caption{WER Results on the Librispeech test data}
   \label{tab:results1}
 \centering
\rowcolors{1}{white!25}{white!50}
\resizebox{5cm}{!}{
\begin{tabular}{ *6l }    \toprule
Model &Clean&Other &  Size (M) \\\midrule
RNN-T &5.9& 15.71&30\\ 
Conformer &5.7&14.24 &29\\ 
ContextNet & 6.02&14.42&28\\
ConvRNN-T & 5.11&13.82&29\\\bottomrule
 \hline
\end{tabular}}
\end{table}

First, we  build an RNN-T whose acoustic encoder and label encoders consist of seven and one  layers of uniLSTM, respectively, with 640 hidden units and dropout of 0.1;  we add  $L_2$ regularization of $1e-6$ to all trainable weights. We use a projection layer after each LSTM layer with a Swish activation. For all other models we keep the label encoder the same as the  RNN-T and only replace the LSTM acoustic encoder with the ContextNet, Conformer, and ConvRNN-T encoders. The dimension of the encoder output  is  set to 512 for all models.  We build a Conformer  with 14 layers in which we use multi-head attention with 4 heads and  each  head with dimension of 64. We make Conformer  causal by applying masks to multi-head attention layers to only attend to left context as well as building all convolutional layers to be causal.  For the convolutional sub-sampling block of Conformer, we use two layers of 2D CNN with  filters of 128 channels, kernel size of three, and stride of two.  The feedforward  hidden unit dimension is set to 1,024.  We build ContextNet in accordance to \cite{han2020contextnet}; we stack 23 blocks, each contains 5 CNN layers except for the first and last blocks. We set all strides to 1 and  filters with 256 and 512 channels  for the first 15  and last 8 blocks, respectively.  We make our ContextNet causal by using 1-D causal CNN to only perform convolution on left frames.  We add more regularization and robustness using SpecAug \cite{park2019specaugment} with the following hyper-parameters: maximum ratio of masked time frames=0.04,  adaptive multiplicity= 0.04, maximum ratio of masked frequencies= 0.34, and number of frequency masks=2.
 We use a word-piece tokenizer and generate 2,500  word-piece tokens as the output vocabulary.   
The number and type of parameters of  ConvRNN-T are given in Table 3 and 4, respectively.
We use  Adam optimizer with $\beta_1$ = 0.9, $\beta_2$= 0.98, and $\epsilon$ = 1e-9.  The learning curve was chosen have high pick of 0.002 and warm-up rate  of 10,000. We used step size of 1000, and 5000 for Librispeech and in-house data, respectively, and  the model trained until no improvement  was observed on the dev set. The  models were trained using four machines each of which has eight  NVIDIA Tesla V100 GPUs.

\begin{table}[t] 
  \caption{Relative WER Reduction on de-identified in-house data}
   \label{tab:results2}
 \centering
\rowcolors{1}{white!25}{white!50}
\resizebox{6cm}{!}{
\begin{tabular}{ *6l }    \toprule
Model &Dev&Test&  Size (M) \\\midrule
RNN-T  & 0.00(ref)&0.00 (ref)&30\\ 
Conformer &+10.70 & +8.30&29\\ 
ContextNet & -8.00&-7.10&28\\
ConvRNN-T & +12.80& +10.10&29\\\bottomrule
 \hline
\end{tabular}}
\end{table}

\subsection{Results}
Table 1 shows the WER obtained from training and testing the models using Librispeech data.  We observe a significant  WER reduction (about 2.11$\%$) compared to RNN-T  for test-other data.  In addition, the results show that ConvRNN-T outperforms both Conformer and ContextNet for both  test-clean and test-other  sets. The observation that ConvRNN-T improvements are more pronounced for the test-other data indicates that our convolutional frontend provides more robust context to LSTM layers when data is more noisy. This improvement is  consistent with the successful usage of convolutional models  for speech enhancement as seen in \cite{pandey2019tcnn}.  The training of all models converge before 100 epochs. The results reported here are obtained using greedy search and no language model is included. 

Table 2 shows the relative word error reduction when we train and test the models using in-house data.  Consistent with Librispeech, ConvRNN-T significantly outperforms RNN-T and 
ContextNet and marginally gives  the same performance as Conformer. 
\begin{table}[t]
  \caption{The number of parameters of ConvRNN-T modules (in million) }
   \label{tab:results1}
 \centering
\rowcolors{1}{white!25}{white!50}
\resizebox{4cm}{!}{
\begin{tabular}{ *6l }    \toprule
Module &Size (M)\\\midrule

Convolution blocks&5.40\\ 
LSTM encoder& 18.93\\
Joint network& 1.28\\
Decoder input embedding& 0.62\\
LSTM decoder & 2.62\\\bottomrule
 \hline
\end{tabular}}
\end{table}
\begin{table}[t]
  \caption{Global encoder model hyper-parameters as shown in Figure 3; we stack 6 blocks. $i$ indicates the block number; $D$ is the input dimension.}
   \label{tab:results4}
 \centering
\rowcolors{1}{white!25}{white!50}
\resizebox{6cm}{!}{
\begin{tabular}{ *5l }    \toprule
Encoder &Channels&Kernel size& Stride& Dilation  \\\midrule
PW CNN  &  2$\times D$ & 1&1&1\\
 DW CNN&  $D$ &3 &1&$2^i$ \\
 PW CNN & $D$& 1& 1&1 \\\bottomrule
 \hline
\end{tabular}}
\end{table}

\begin{table}[t]
\small
  \caption{The impact of the local and global CNN encoders on ConvRNN-T performance. The numbers are relative WER reduction. The number of parameters are in millions.}
   \label{tab:fig7}
 \centering
\rowcolors{1}{white!25}{white!50}
\resizebox{6cm}{!}{
\begin{tabular}{ *5l }    \toprule
Local& Global&Dev&Test & Size (M)  \\\midrule
\cmark &\xmark& 0.00 (ref)& 0.00 (ref)&2.9\\
\xmark&\cmark& -35.80 &-40.21 &2.5\\
 \cmark& \cmark &+9.40 &+9.01 &5.4  \\\bottomrule
\end{tabular}}
\end{table}

\begin{figure}[h]
  \centering
  \includegraphics[width=.9\linewidth]{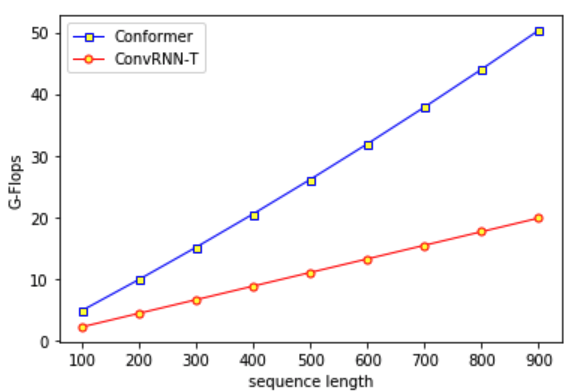}
  \caption{ The number of flops against the length of sequence (number of frames) for ConvRNN-t and Conformer.}
  \label{fig:flops}
\end{figure}
 \subsection{Impact of the global  and local CNN encoders on performance}
In order to investigate the impact of the local and global CNN encoders individually on WER reduction, we build two ConvRNN-Ts: in the first one, we exclude the global encoder and in the second one we exclude the local encoder from the architecture shown in Figure 1. Table 5 shows the relative WER reduction obtained from these models trained using our in-house de-identified data. As indicated in the table,  inserting the global encoder on top of the local encoder reduces WER by 9$\%$ relative. The results also indicate the existence of the local encoder in ConvRNN-T is necessary and using solely the global encoder degrades the performance significantly. 
\subsection{Computational complexity}
Generally, computational complexity of  LSTM, CNN, and multi-head attention (MHA) are of order of $O(n d^2)$, $O(kn d^2)$, and $O(n^2  d)$, respectively \cite{vaswani2017attention}, where $n$, $d$, and $k$ denote the length of sequence, feature dimension, and kernel size,  respectively.  In order to obtain a relative approximation of computational cost for ConvRNN-T vs Conformer, we  calculate the number of flops executed in their encoders. The number of flops for a convolutional layer is approximately computed as product  of $4 \times C_i \times k^2 \times Co \times W \times H$ , where the parameters, respectively,  are the number of input channels, kernel size, the number of output channels,  sequence length and feature dimension.  The number of flops for a stack of $l$ LSTM layers computed by $8 \times l \times s \times (I+d)\times d$, where $s$, $d$, and $I$ are the sequence length, hidden size, and feature dimension \cite{flopslstmmin} . We  compute the number of flops for MHA and the feedforward net in the Conformer encoder using the Electra package \cite{clark2020electra} and add them with the number of CNN FLOPS used in Conformer. Figure 4 compares GFLOPs ( 1GFLOP = one billion  floating-point operations) for two models for sequences with different lengths. As shown in the figure, ConvRNN-T performs less GFLOPs compared to Conformer and as the length of sequence increases the gap becomes wider as expected.
It should be  noted recently several approaches  are proposed to reduce the computational complexity of MHA \cite{choromanski2020rethinking,wang2020linformer,li2021efficient}. In these models the cost reduction is, however, noticeable  only for very long sequences and/or  comes with some accuracy degradation tradeoff.
 \section{Conclusion}
In this paper, we introduced ConvRNN-T, a new streaming E2E ASR model that significantly improves the performance of  RNN-T and outperforms SOTA models.  ConvRNN-T consists of local and global convolutional context based encoders that provide both local and global context to LSTM layers. Our work signifies the importance of convolutional neural networks as frontend context encoders for audio signals. We showed the computational complexity of ConvRNN-T is less than that of the best Transformer-based ASR model, i.e. Conformer. ConvRNN-T is suitable for streaming on-device applications where we want to take advantage of both CNNs and LSTMs to have  accurate, fast, yet small ASR footprint.  In future works, we would apply our  CNN based frontend to multi-channel streaming ASR.

\bibliographystyle{IEEEtran}

\bibliography{mybib}

\end{document}